\documentclass[printer]{aa}
\usepackage{psfig}

\def\ltsima{$\; \buildrel < \over \sim \;$}
\def\lsim{\lower.5ex\hbox{\ltsima}}
\def\gtsima{$\; \buildrel > \over \sim \;$}
\def\gsim{\lower.5ex\hbox{\gtsima}}

\begin{document}

\title{The afterglow of GRB~021004: surfing on density waves}
\titlerunning{GRB~021004}

\author{Davide Lazzati\inst{1}, Elena Rossi\inst{1}, Stefano Covino\inst{2},
Gabriele Ghisellini\inst{2} \and Daniele Malesani\inst{2}}
\authorrunning{Lazzati et al.}

\institute{Institute of Astronomy, University of Cambridge, Madingley
Road, CB3 0HA Cambridge, England \\
\email{lazzati@ast.cam.ac.uk}
\and INAF -- Osservatorio Astronomico di Brera, via Bianchi 46, I-23807 
Merate, Italy}

\date{Received 15 Oct. 2002}

\abstract{
We present a model for the early optical afterglow of GRB~021004.
This burst had one of the earliest detected optical afterglows,
allowing for a dense optical sampling.  The lightcurve was peculiar,
since bright bumps were superimposed to the regular power-law decay
observed in many other events.  We show that, given their time scale
and shape, the bumps are likely due to the interaction of the fireball
with moderate density enhancements in the ambient medium.  The
enhancements have a density contrast of order 10, modifying only
slightly the dynamics of the fireball, which therefore surfs on them
rather than colliding into them.  A relativistic reverse shock does
not develop.  Since the interaction takes place when the fireball is
still hyper-relativistic it is not possible to understand if the
overdensities are localized in clumps or are spherically symmetric
around the GRB progenitor.  The monotonic decrease of the contrast of
successive rebrightenings suggests however the presence of clumps
embedded in a uniform environment. Such an interpretation,
complemented by the detection of several high velocity absorption
systems in the optical spectrum, strongly suggests that GRB~021004
exploded within the remnant of a previous explosion.
\keywords{Gamma rays: bursts -- Radiation mechanisms: non-thermal --
ISM: structure} }

\maketitle

\section{Introduction}
\label{sec:int}
On the 4th of October 2002 at 12:06:13 UT, HETE II detected a burst of
$\sim$100 s duration (Shirasaki et al. 2002) with a 7--400~keV fluence
of $3.2\times10^{-5}$ erg cm$^{-2}$ (Lamb et al. 2002).  The fast
accurate localization allowed the robotic telescope Oschin/NEAT to
detect the optical afterglow (OA) $\sim$9 minutes after the trigger
(Fox 2002), at the level of $R=15.52$.  Observations performed earlier
by Torii et al. (2002) yielded upper limits around $R\sim 13.6$,
$\sim$3.5 minutes after the trigger.  The prompt OA identification
allowed a very dense sampling of its light curve at early times (see
the references listed in the caption of Fig. 1), and spectroscopic
observations at medium to high resolution (Matheson et al. 2002;
M{\o}ller et al. 2002; Fox et al. 2002; Anupama et al. 2002; Eracleous et
al. 2002; Chornock \& Filippenko 2002; Mirabal et al. 2002b; Sahu et
al. 2002c; Salamanca et al. 2002; Djorgovski et al. 2002; Savaglio et
al. 2002).  The spectra revealed an emission line interpreted as
Ly-$\alpha$ at $z=2.328$ and several absorption lines at slightly
different redshifts, corresponding to velocity differences up to
3000--4000 km s$^{-1}$, suggesting that the absorbing material could
be either a clumpy wind ejected by a massive star progenitor or a
clumpy remnant of a precursor supernova explosion (Salamanca et
al. 2002; Mirabal et al.  2002b).  The X-ray afterglow (XA), observed
by {\it Chandra} (Sako \& Harrison 2002) decays in time as $t^{-1\pm
0.2}$ with an average flux of $4.3\times 10^{-13}$ erg cm$^{-2}$
s$^{-1}$. A preliminary spectral analysis shows no distinct features.

\begin{figure*}
\psfig{file=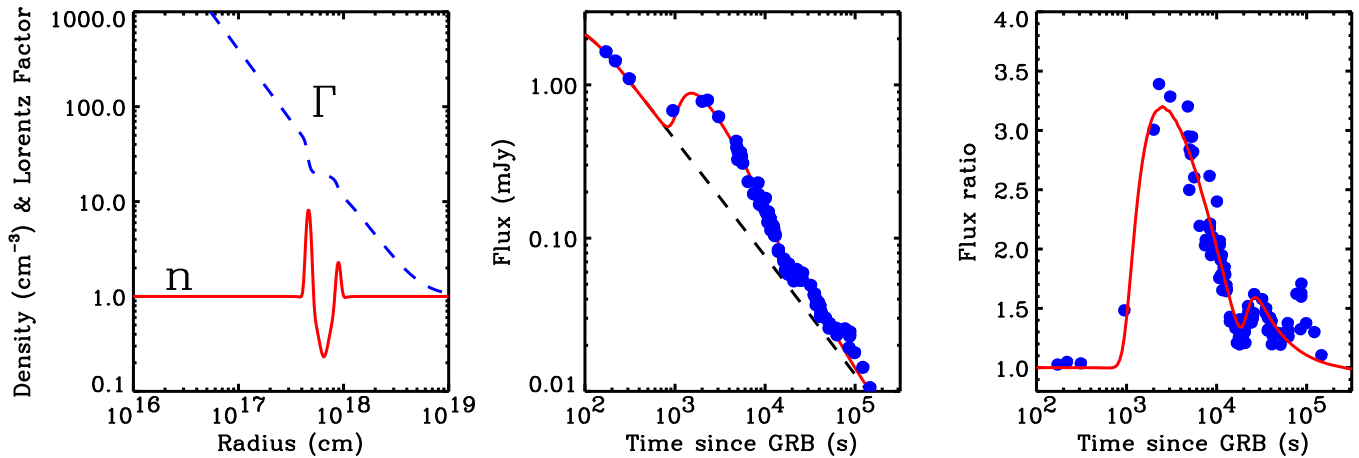,width=1.\textwidth}
\caption{{ISM model and fits to the $R$-band light curve of GRB~021004.
The left panel shows the density model, with an enhancement at
$R\sim5\times10^{17}$~cm. The dashed line shows the evolution of the
Lorentz factor of the fireball. The central panel shows a compilation
of R-band photometry from GCNs. The solid line shows the lightcurve
predicted for the density structure shown in the left panel, while the
dashed line shows how the lightcurve would be in a uniform ISM. The
right panel shows an emphasized version of the central one, in which
the lightcurve variability is enhanced by dividing the data and the
model by the uniform ISM case (the dashed line in the central
panel). A third bump may be present at $t\sim10^5$~s, but we did not
attempt to model it given the paucity of the data. In addition,
further complications should be considered, since our model predicts
the transit of the cooling break at $t\sim2$~d (possibly observed by
Matheson et al. 2002), and a jet break is expected to appear at
$1.5\lsim{t}\lsim10$~d (Malesani et al. 2002). All the times are in
the rest frame of the GRB host, and no correction for reddening was
applied.  Magnitudes are calculated using the calibration of Henden
2002a.  Data from: Anupama et al. 2002; Balman et al. 2002; Barsukova
et al. 2002; Bersier et al. 2002; Cool \& Schaefer 2002; Di Paola et
al. 2002; Fox 2002; Halpern et al. 2002a, 2002b; Holland et al. 2002b,
2002c; Klotz \& Boer 2002; Klotz et al. 2002; Malesani et al. 2002;
Masetti et al. 2002; Matsumoto et al. 2002a, 2002b; Mirabal et
al. 2002a, 2002b; Oksanen \& Aho 2002; Oksanen et al. 2002; Sahu et
al. 2002a, 2002b; Stanek et al. 2002; Stefanon et al., 2002; Uemura et
al. 2002; Weidinger et al. 2002; Winn et al. 2002; Zharikov et
al. 2002.  }
\label{fig1}}
\end{figure*}

The peculiarity of this OA is the presence of a major rebrightening,
with a rise time $t_{\rm{rise}}\sim t_{\rm{start}}$ where
$t_{\rm{start}}$ is the moment in which the rebrightening starts (see
the central panel of Fig.~\ref{fig1}). The flux then reconnects to the
extrapolation of the early time observations for $15\lsim{t}\lsim100$
hours after the GRB. In this time span, the lightcurve shows at least
one (maybe two) additional bumps, but with smaller contrast. Such a
behavior is unprecedented and is not due to a calibration problem
(Henden 2002b).

In the lightcurve of GRB~970508, a major rebrightening was observed
from the optical to the X-rays (Galama et al. 1998; Piro et al. 1998);
in this case, however, after the rebrightening the flux remained
larger than the extrapolation of the early time data and no additional
feature was detected for a long time.  Such a behavior was interpreted
as due to the impact of a late shell of the fireball that gave
additional energy to the forward shock (Panaitescu et al. 1998). In
the case of GRB~000301C an achromatic rebrightening was observed. It
had a very small time scale compared to the time at which it took
place and was consequently interpreted as due to a micro-lensing
effect (Loeb \& Perna 1998; Garnavich et al. 2000).

In this Letter we propose that the rebrightenings observed in the
lightcurve of GRB~021004 are the result of the interaction of the
fireball with density enhancements (and possibly gaps) in the
surrounding medium (Wang \& Loeb 2000; Dai \& Lu 2002).  These
inhomogeneities may either be in the form of spherical shells --
possibly due to instabilities in the wind ejection history of the
progenitor star -- or of clumps and filaments -- possibly associated
to a cooling supernova remnant (SNR). This latter interpretation seems
to be corroborated by multiwavelength photometry and spectroscopy.

\section{The model}

We applied a numerical code to compute the lightcurve. The code takes
fully into account the light travel-time effects, and assumes standard
synchrotron emission equations to compute the comoving frame intensity
(Panaitescu \& Kumar 2000; Granot \& Sari 2002). The fireball dynamics
in a non uniform medium is treated by a numerical integration of the
energy and momentum conservation equations (Rhoads 1999; Kumar \&
Panaitescu 2000), with the possibility of adding a reverse shock if a
large density contrast is present.

\begin{figure*}
\psfig{file=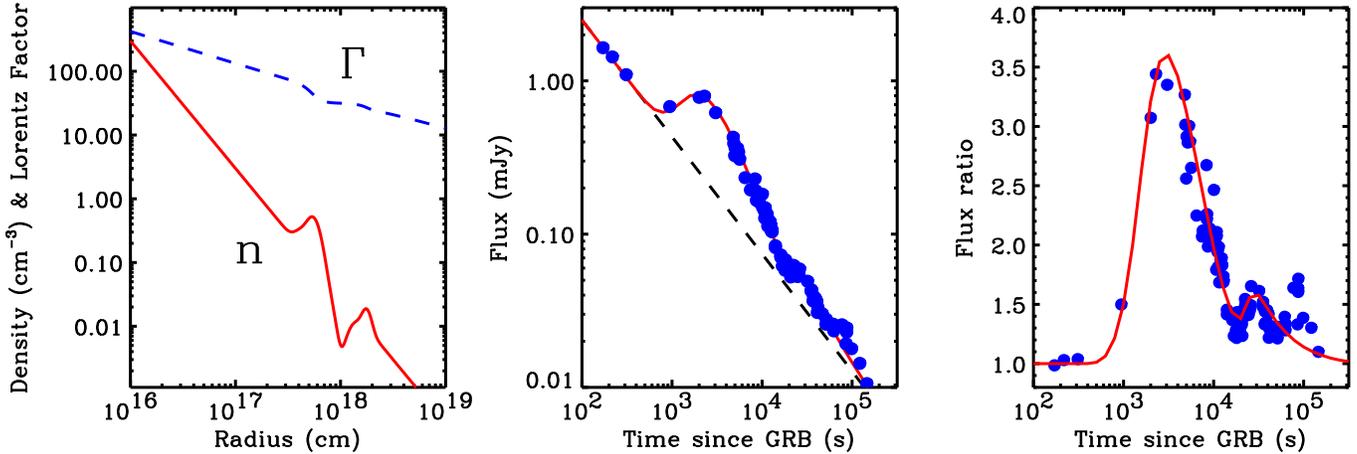,width=1.\textwidth}
\caption{{Same as Fig.~\ref{fig1} but for a wind environment.}
\label{fig2}}
\end{figure*}

The code assumes spherical symmetry for the environment properties and
therefore we can model only radial inhomogeneities in the external
medium. The results can be extended to a clumpy geometry with simple
considerations (see \S~3). We did not include the detailed treatment
of the fireball impact (Dai \& Lu 2002) since the transient features
that are predicted have a time scale much smaller than that due to the
curvature of the fireball and are therefore smeared out by the
integration on the equal arrival time surface.

When the fireball impacts on an overdensity, it goes through a
transient phase, in which the flux increases sharply, relaxing
asymptotically to the solution for an unperturbed medium with the
higher density (Sari \& Piran 1995). The effect of the interaction
depends on the spectral range, on the radial dependence of the average
density (wind or interstellar medium, hereafter ISM) and on the
cooling regime of the electrons. Since the transitory increase is
smeared out, the increase in the flux can be estimated with the
asymptotic solution. In a uniform density ISM for the slow cooling
electron regime, an observation at a frequency above the cooling break
will be insensitive to the density variation, while an observation
between the peak (in $F_\nu$ vs $\nu$) and the cooling frequency will
yield a flux increase $F_1/F_0\propto(n_1/n_0)^{1/2}$, where the
subscripts $_0$ and $_1$ refer to the smaller and larger densities,
respectively.  In a wind environment, the same observation would
depend linearly on the density contrast. In the case of fast cooling
electrons any observation above the peak frequency will be insensitive
on the density, both for the ISM and wind environments. In the radio
band, the behavior can be even more complicated, since below the
self-absorption break the flux will decrease in response to a density
increase, while above the self-absorption the flux will
increase. Finally, the situation can be made even more complicated by
the fact that the break frequencies are shifted by the overdensity and
may cross the observational band during the rebrightening.

As mentioned above, if the fireball bumps up against a density jump, a
reflected (as well as a forward) shock can be generated, which
propagates back into the hot relativistic shell.  Its capability of
modifying the fireball dynamics depends on the density contrast
$n_1/n_0$.  If $n_1/n_0>250$ the bulk Lorentz factor of the reverse
shocked hot shell is significantly ($\sim0.3$) lower than that of the
unshocked hot shell (at a fixed radius), since the relative Lorentz
factor of the two shells is $\Gamma_{\rm{rel}}>2$.  Therefore such a
relativistic reverse shock substantially slows down the incoming
fireball, converting most of the initial kinetic energy into internal
energy. Its contribution to the emission must be taken into account.
This is not the case with GRB~021004: the fireball encounters smaller
density contrasts (see below and the left panels of Figs.~\ref{fig1}
and~\ref{fig2}) and the hot shell continues its run with an
asymptotically almost unchanged Lorentz factor.  Therefore in the
following we will neglect any contribution from the reverse shock
emission.

\section{Results}

Figures~\ref{fig1} and~\ref{fig2} show our model in the cases of
an ISM and wind environment, respectively. Let us first discuss the
wind case in Fig.~\ref{fig2}, which we consider less likely.

As discussed above, a requirement in order to observe a rebrightening
in a given band ($R$ in our case) is that the electrons must be in the
slow cooling regime and the band must lie between the peak and cooling
frequencies. Such a constraint is easily fulfilled in a wind
environment, since the cooling frequency increases with time. Under
these conditions, the lightcurve should decay as a power law with an
index\footnote{This relation holds if $p<2$ (Dai \& Cheng 2001). We
use it since the $p>2$ relation $\delta_w=(3p-1)/4$ yields
inconsistently $p=1.35$.} $\delta_w = (p+8)/8$ where $p$ is the
power-law index of the electron distribution
[$n(\gamma)\propto{}\gamma^{-p}$] and $\delta_w$ is defined through
$F(t)\propto{}t^{-\delta}$. The fitted lightcurve decay is
$\delta\sim0.75$, which would imply an unphysical distribution with
$p<0$.  In addition (see Fig.~\ref{fig2}) the density enhancements
take place at a fairly large distance from the progenitor star, when
the wind density is very small ($n\sim0.3$~cm$^{-3}$). In these
conditions, a more complicated density structure, due to the
interaction of the wind with the ISM, would be expected (Ramirez-Ruiz
et al. 2001).

We consider the case of a uniform environment more likely. In this
case, the relation between the temporal decay of the OA and the
electron distribution is $\delta_{\rm{ISM}}=3(p-1)/4$, yielding
$p\approx2$. This value is theoretically acceptable, and is comparable
to the values derived in other afterglows (Panaitescu \& Kumar
2002). Moreover, it is consistent with the X-ray spectrum and temporal
decay detected by {\it Chandra} at later times (Sako~\& Harrison
2002), if the X-ray band lies above the cooling frequency. Again, this
is usual in observed XAs (see e.g. GRB~010222, in't Zand et al. 2002).
The required density contrast to explain the bump is:
$n_1/n_0\approx8.5$ (Fig.~\ref{fig1}).

The density of the ISM and the other afterglow parameters can be
constrained by requiring that the optical $R$ band lies between the
peak and cooling frequencies and that the OA flux is consistent with
observations. Such a procedure yields a moderately low density,
$n\sim1$~cm$^{-3}$, of the uniform part of the environment.  For the
fit shown in Fig.~\ref{fig1} we also assumed an electron energy
fraction $\epsilon_e=0.01$, a magnetic field energy fraction
$\epsilon_B=0.001$, and an efficiency of converting the kinetic energy
of the fireball into photons during the prompt phase of $\eta=5\%$.
It should be emphasized, however, that modeling with the standard
afterglow theory is sensitive to some simplistic assumptions about the
shock physics and about the magnetic field generation mechanisms (see,
e.g., Rossi \& Rees 2002). Due to the lack of a complete broad-band
coverage of the lightcurve, a larger density could be envisaged. A
robust upper limit to the average density can however be set by
considering that a non-relativistic transition was not observed
several days after the explosion. This yields
$n\le10^{6}$~cm$^{-3}$. It should also be mentioned that, even though
all the magnitudes are corrected for the comparison star discussed by
Henden (2002b), the data we are attempting to model have been taken
with different telescopes and therefore absolute values and errors
should be taken with caution. In addition, significant fast variability
was detected (Halpern et al. 2002a). For these reasons we did not
attempt to perform a formal fit to the data set, but rather to
reproduce its general behavior. We also did not try to model the very
small time scale variability of the lightcurve, which may be produced
by small scale ISM turbulence as in GRB~011211 (Holland et al. 2002a).

What can be safely concluded from the observed rise time is that the
overdensity lays at a distance
$R\sim5\times10^{17}\,(E_{54}/n)^{1/4}$~cm from the explosion centre,
a value that is fairly independent of the assumed
density. Interestingly, the value of the Lorentz factor at the
beginning of the interaction is even more robustly constrained, being
$\Gamma_1\sim50\,(E_{54}/n)^{1/8}$. This, together with the fact that
we can reproduce the bump shape with a spherical overdensity (we do
not see the edges of the clump), allows us to put a lower limit to the
angular size of the clump $\theta_1\gsim{}1^\circ$. Interestingly, the
Gaussian density enhancement we used has a radial width
$\delta{}R/R=0.04$, comparable to the inferred lower limit on the
angular size. Two more considerations support the idea that the
density structure is indeed clumpy rather than made by under and
over-dense shells. First, we had to include in the radial density
structure an under-dense part, due to the need of reproducing at best
the decaying part of the first enhancement. The same behavior can be
due to a clump that has an angular size similar to the relativistic
beaming of the fireball ($\theta\sim1/\Gamma$). As soon as the
fireball is slowed down by the interaction with the clump, the edges
of the clump can be detected, with a corresponding decrease of
flux. Secondly, the luminosity ratio of the second (and possibly the
third) bump with respect to the underlying power-law is smaller than
the contrast of the first bump (see right panel of
Fig.~\ref{fig1}). Our radial overdensity had therefore to be smaller
(see Fig.~\ref{fig1}). However, should a clump with the same
overdensity and size of the first one be present at a larger distance,
it would produce a bump in the lightcurve with a flux increase of a
factor ${\cal{}R}_2\approx1+{\cal{R}}_1\,(\Gamma_2/\Gamma_1)^2$, where
${\cal{R}}_1$ is the flux ratio for the main bump and
$\Gamma_2<\Gamma_1$ is the Lorentz factor of the fireball at the
moment of the interaction with the second clump. This is due to the
fact that the second bump would interact with a smaller portion of the
visible area of the fireball.  The predicted ratio is
$(\Gamma_1/\Gamma_2)^2\approx7$ (left panel of Fig.~\ref{fig1}), in
excellent agreement with the data (right panel of Fig.~\ref{fig1}). As
a consequence, the lightcurve should evolve to a smooth decay due to
(i) the smaller area of the visible fireball that would be affected by
the interaction with a clump and (ii) the possibility of interaction
with more and more clumps simultaneously.

We therefore conclude that the most likely environment for GRB~021004
is a uniform medium with clumps of density contrast of order 10 and
size $\Delta{R}\sim10^{16}$~cm. Such an environment may be quite
typical for relatively young SNRs (B\"ottcher et al. 2002). This
evidence adds to the multiple high velocity absorption systems
detected in the optical spectra (Salamanca et al. 2002), suggesting
that the explosion of the burst took place within the remnant of a
former explosion, likely a SN.

\section{Discussion and conclusions}

We have presented a model to fit the OA of GRB~021004, which is
characterized by the presence of bright bumps on top of the usual
power-law decay. We suggest that the most likely origin for the bumps
is the interaction of the fireball with density clumps, with a density
contrast $n_1/n_2\sim10$ and size $\Delta{R}\sim10^{16}$~cm. These
clumps should lie at a distance of the order of
$\sim5\times10^{17}$~cm from the explosion site, even though this
number is uncertain due to the lack of a precise measure of the
average density of the environment.

One possible alternative to this explanation is the presence of a very
narrow and energetic component of the fireball, lying slightly
off-axis from the line of sight (Panaietscu et al. 1998). In this case
such a bump should be a common feature of all OAs that an
unprecedented early monitoring has now disclosed.  Such a component,
was however not seen in GRB~990123. In addition, the equality of
$t_{\rm{start}}$ with the rise time $t_{\rm{rise}}$ would be a mere
coincidence. It would also be difficult to explain the second (and
possible third) rebrightening. Alternatively, the bump may be produced
by a neutron decay trail ahead of the external shock (Beloborodov
2002). Again, a single bump would be expected in the simplest case.

Evidence for a clumpy geometry of the medium surrounding the GRB
explosion site comes also from the detection, in the optical spectra,
of multiple absorption features from intermediate ionization ions
(e.g. CIV). The large velocity spacing of $\sim3000$~km/s, suggests a
clumpy medium outflowing from the explosion site (Mirabal et al. 2002;
Salamanca et al. 2002). The possible physical association of the
absorbing clouds with those producing the afterglow rebrightenings is
tantalizing, even though the former should lie at a larger distance,
since all the carbon atoms are completely ionized within a distance
$R\gsim10^{19}$~cm from the explosion site (a density of
$\sim10^{7}$~cm$^{-3}$ would be required to have recombination of free
electrons onto the carbon nuclei in $\sim1$~day).

We speculate that the burst exploded within a Crab-like SNR, produced
by a supernova that exploded $10-100$ years before the GRB. Such a
time interval is in the upper extreme, yet consistent, with what
predicted by the Supranova scenario (Vietri \& Stella 1998). This
possibility was also suggested to account for the possible detections
of X-ray lines (Piro et al. 1999, 2000; Antonelli et al. 2000; Reeves
et al. 2002) in the early XA of several GRBs. In those cases, the
SN-GRB delay should have been smaller (Lazzati et al. 1999,
2002). Consistently, no X-ray feature is detected in the X-ray
spectrum of GRB~021004 (Sako \& Harrison 2002) since the remnant is
too far from the explosion site and its density too small. Also, the
outflow velocity of the SNR inferred for this burst is much smaller
than that required in bursts with X-ray features, consistent with the
slow-down of the SN ejecta with time. Finally, if this interpretation
is correct, the OA of GRB~021004 should not have a SN component in its
lightcurve (see, e.g., Bloom et al. 2002), even though such a
component would be in any case difficult to detect given the large
redshift of the event.

\begin{acknowledgements}
We thank Martin Rees for useful discussions and for critically reading
the manuscript.  DL and ER acknowledge financial support from the
PPARC and the Isaac Newton Fellowship, respectively.
\end{acknowledgements}


\begin{thebibliography}{}
\bibitem{} Antonelli L.~A., Piro L., Vietri M., et al., 2000, ApJ,  545, L39
\bibitem{} Anupama G.C., Sahu D.K., Bhatt B.C. \& Prabhu T.P., 2002, GCN 1582
\bibitem{} Balman S., Esenoglu H., Parmaksizoglu M., Aslan Z. \& 
    Kiziloglu U., 2002, GCN 1580
\bibitem{} Barsukova E.A., Goranskij V.P., Beskin G.M., 
      Plokhotnichenko V.L. \& Pozanenko A.S., 2002, GCN 1606
\bibitem{} Beloborodov A., 2002, ApJ submitted (astro-ph/0209228)
\bibitem{} Bersier D., Winn J., Stanek K.Z. \& Garnavich P., 2002, GCN 1586
\bibitem{} Bloom J.~S., Kulkarni S.~R., Price P.~A., et al., 2002, 
	ApJ,  572, L45
\bibitem{} B{\" o}ttcher M., Fryer C.~L., Dermer C.~D., 2002, ApJ,  567, 441
\bibitem{} Chornock R. \& Filippenko V., 2002, GCN 1605
\bibitem{} Cool R.J. \& Schaefer J.J., 2002, GCN 1584 
\bibitem{} Dai Z.~G., Cheng K.~S., 2001, ApJ,  558, L109
\bibitem{} Dai Z.~G., Lu T., 2002, ApJ,  565, L87
\bibitem{} Di Paola A., Boattini A., Del Principe M., 
      et al. 2002, GCN 1616 
\bibitem{} Djorgovski S.G., Barth A., Price P. et al., 2002, GCN 1620
\bibitem{} Eracleous M., Schaefer B.E., Mader J., Wheeler C., 2002, GCN 1579
\bibitem{} Fox D.W., 2002, GCN 1564
\bibitem{} Fox D.W., Barth A.J., Soderberg A.M. et al., 2002, GCN 1569 
\bibitem{} Galama T.~J., Groot P.~J., van Paradijs J., et al., 1998, ApJ,  
	497, L13
\bibitem{} Garnavich P.~M., Loeb A., Stanek K.~Z., 2000, ApJ,  544, L11
\bibitem{} Granot J., Sari R., 2002, ApJ,  568, 820
\bibitem{} Halpern J.P., Armstrong E.K., Espailla C. C. \& Kemp J.,  
     2002a, GCN 1578
\bibitem{} Halpern J.P., Mirabal N., Armstrong E.K., Espaillat C. C. 
     \& Kemp J.,  2002b, GCN 1593
\bibitem{} Henden A., 2002a, GCN 1583 
\bibitem{} Henden A., 2002b, GCN 1630
\bibitem{} Holland S.~T., Soszy{\'n}ski I., Gladders M.~D., et al., 
	2002a, AJ,  124, 639
\bibitem{} Holland S.~T., Fynbo J.P.U., Weidinger M., Egholm M.P. \& Levan A., 
     2002b, GCN 1585
\bibitem{} Holland S. T., Fynbo J.P.U., Weidinger M. et al. 2002c, GCN 1597
\bibitem{} in't Zand J.J.M., Kuiper L., Amati L., et al. 2002, ApJ, 559, 710
\bibitem{} Klotz A. \& Boer M., 2002, GCN 1614
\bibitem{} Klotz A., Boer M. \& Thuillot W.,  2002, GCN 1615
\bibitem{} Kumar P., Panaitescu A., 2000, ApJ,  541, L9
\bibitem{} Lamb D., Ricker G., Atteia J--L. et al., 2002, GCN 1600
\bibitem{} Lazzati D., Campana S., Ghisellini G., 1999, MNRAS, 304, L31
\bibitem{} Lazzati D., Ramirez-Ruiz E., Rees M.~J., 2002, ApJ,  572, L57
\bibitem{} Loeb A., Perna R., 1998, ApJ,  495, 597
\bibitem{} Malesani D., Covino S., Ghisellini G., et al., 2002, GCN 1607 
\bibitem{} Masetti N., Pizzichini G., Bartolini C. et al., 2002, GCN 1603
\bibitem{} Matheson T., Garnavich P.M., Foltz C. et al., 2002, 
	ApJL submitted (astro-ph/0210403)
\bibitem{} Matsumoto K., Kawabata T., Ayani K., Urata Y. 
       \& Yamaoka H., 2002a, GCN 1567
\bibitem{} Matsumoto K., Kawabata T., Ayani K., et al. 2002b, GCN 1594    
\bibitem{} Mirabal N., Armstrong E.K., Halpern J.P. \& Kemp J., 
     2002a, GCN 1602
\bibitem{} Mirabal N., Halpern J.P., Chornock P. \& Filippenko A.V., 
     2002b, GCN 1618
\bibitem{} M{\o}ller P., Fynbo J. P. U., Hjorth J., et al., 2002,
	A\&A in press (astro-ph/0210654)
\bibitem{} Oksanen A. \& Aho M., 2002, GCN 1570 
\bibitem{} Oksanen A., Aho M., Rivich K., et al. 2002, GCN 1591
\bibitem{} Panaitescu A., Meszaros P., Rees M.~J., 1998, ApJ,  503, 314
\bibitem{} Panaitescu A., Kumar P., 2000, ApJ,  543, 66
\bibitem{} Panaitescu A., Kumar P., 2002, ApJ,  571, 779
\bibitem{} Piro L., Amati L., Antonelli L.~A., et al., 1998, A\&A,  331, L41
\bibitem{} Piro L., Garmire G., Garcia M., et al., 2000, Science,  290, 955 
\bibitem{} Piro L., Costa E., Feroci M., et al., 1999, ApJ,  514, L73
\bibitem{} Ramirez-Ruiz E., Dray L.~M., Madau P., Tout C.~A., 2001, 
	MNRAS, 327, 829
\bibitem{} Reeves J.~N., Watson D., Osborne J.~P., et al., 2002, Nat, 416, 512
\bibitem{} Rhoads J.~E., 1999, ApJ,  525, 737
\bibitem{} Rossi E., Rees M. J., 2002, MNRAS in press (astro-ph/0204406)
\bibitem{} Sahu D.K., Bhatt B.C. \& Prabhu T.P., 2002a, GCN 1581
\bibitem{} Sahu D.K., Bhatt B.C., Anupama G.C. \& Prabhu T.P., 2002b, GCN 1587
\bibitem{} Sahu D.K., Fruchter A., Burud I. \& Sembach K., 2002c, GCN 1608
\bibitem{} Sako M. \& Harrison F.A., 2002, GCN 1624 
\bibitem{} Salamanca I., Rol E., Wijers R., et al. 2002, GCN 1611
\bibitem{} Sari R., Piran T., 1995, ApJ,  455, L143
\bibitem{} Savaglio S., Fiore F., Israel G.L. et al., 2002, GCN 1633
\bibitem{} Shirasaki Y., Graziani C. Matsuoka M. et al., 2002, GCN 1565
\bibitem{} Stanek K.Z., Bersier D. \& Winn J., 2002, GCN 1598
\bibitem{} Stefanon M., Covino S., Malesani D. et al., 2002, GCN 1623 
\bibitem{} Torii K., Kato T. \& Yamaoka H., 2002, GCN 1589 
\bibitem{} Uemura M., Ishioka R., Kato T. \& Yamaoka H., 2002, GCN 1566 
\bibitem{} Vietri M., Stella L., 1998, ApJ,  507, L45
\bibitem{} Wang X., Loeb A., 2000, ApJ,  535, 788
\bibitem{} Weidinger M., Egholm M.P., Fynbo J.P. et al., 2002, GCN 1573
\bibitem{} Winn J., Bersier D., Stanek K.Z., Garnavich P. \& Walker A., 
      2002, GCN 1576
\bibitem{} Zharikov S., Vazquez R. \& Benitez G., \& del Rio S., 2002, GCN 1577


\end{thebibliography}
\end{document}